\documentclass{article}

\usepackage[preprint]{neurips_2022}
\usepackage{microtype}
\usepackage{graphicx}
\usepackage{booktabs} 
\usepackage{amsthm}
\usepackage{bbm}
\usepackage{amsmath}
\usepackage{amssymb}
\usepackage{amsfonts}  
\usepackage{bm}   
\usepackage{xcolor}
\usepackage{stmaryrd}
\DeclareMathOperator*{\bigtimes}{\vartimes}

\newtheorem{theorem}{Theorem}
\newtheorem{lemma}{Lemma}

\usepackage[noend]{algorithmic}
\usepackage{subcaption}
\usepackage{caption}
\usepackage{todonotes}
\usepackage{float}
\usepackage{nicefrac}

\usepackage{hyperref}

\usepackage{algorithm}
\usepackage{wrapfig}

\definecolor{darkgreen}{RGB}{0,125,0}

\title{Self-Play PSRO: Toward Optimal Populations in Two-Player Zero-Sum Games}

\author{%
  Stephen McAleer \\
  Carnegie Mellon University\\
  \texttt{smcaleer@cs.cmu.edu} \\
   \And
  JB Lanier\\
  Department of Computer Science\\
  University of California, Irvine\\
  \texttt{jblanier@uci.edu}
  \And
  Kevin A. Wang\\
  Department of Computer Science\\
  University of California, Irvine\\
  \texttt{kevinwang@kevinwang.us}\\
  \And
  Pierre Baldi\\
  Department of Computer Science\\
  University of California, Irvine\\
  \texttt{pfbaldi@ics.uci.edu}\\
  \And
  Roy Fox\\
  Department of Computer Science\\
  University of California, Irvine\\
  \texttt{royf@uci.edu}\\
  \And
   Tuomas Sandholm \\ 
  Carnegie Mellon University\\
  Strategy Robot, Inc.\\
  Optimized Markets, Inc.\\
  Strategic Machine, Inc.\\
  \texttt{sandholm@cs.cmu.edu} \\
}

\begin{document}

\maketitle

\begin{abstract}
In competitive two-agent environments, deep reinforcement learning (RL) methods based on the \emph{Double Oracle (DO)} algorithm, such as \emph{Policy Space Response Oracles (PSRO)} and \emph{Anytime PSRO (APSRO)}, iteratively add RL best response policies to a population. Eventually, an optimal mixture of these population policies will approximate a Nash equilibrium. However, these methods might need to add all deterministic policies before converging. In this work, we introduce \emph{Self-Play PSRO (SP-PSRO)}, a method that adds an approximately optimal stochastic policy to the population in each iteration. Instead of adding only deterministic best responses to the opponent's least exploitable population mixture, SP-PSRO also learns an approximately optimal stochastic policy and adds it to the population as well. As a result, SP-PSRO empirically tends to converge much faster than APSRO and in many games converges in just a few iterations. 
\end{abstract}

\section{Introduction}
In competitive two-agent environments, also known as zero-sum games, deep reinforcement learning~(RL) methods based on the \emph{Double Oracle (DO)} algorithm \citep{double_oracle}, such as \emph{Policy Space Response Oracles (PSRO)} \citep{psro} are some of the most promising methods for finding approximate Nash equilibria in large games. One reason is that they are simple to use with existing RL methods and naturally provide a measure of approximate exploitability, i.e. performance against the opponent's best response. A second reason is that they effectively prune the game tree by only considering mixtures over policies that are already trained to be best responses. Finally, they can be used in games with large or continuous action spaces because they do not require full game-tree traversals. Methods based on PSRO such as AlphaStar \citep{alphastar} and Pipeline PSRO \citep{mcaleer2020pipeline} have achieved state-of-the-art performance on Starcraft and Stratego, respectively. 

PSRO-based methods iteratively add RL best-response policies to a population. The best response for each player trains against a restricted distribution over the opponent's existing population of policies. To find this restricted distribution, a Nash equilibrium (a pair of mutually best-responding policies) is computed in the restricted single-step game where each action correspond to choosing a policy from the population. Eventually, an optimal distribution over these population policies will approximate a Nash equilibrium in the full game. 

Because PSRO adds pure-strategy (i.e. deterministic) best responses in each iteration, PSRO may need to add many policies to the population before they can support a Nash equilibrium. In fact, in certain games, all pure strategies will be added before finding a Nash equilibrium. This is because many games require mixing over a large number of pure strategies to arrive at a Nash equilibrium. Furthermore, before termination, the restricted distribution over population policies can be arbitrarily exploitable. 

To address this second issue, APSRO~\citep{anytime_psro} learns a restricted distribution that minimizes regret against the opponent's best response policy while the latter is training. The resulting distribution approximates the least-exploitable restricted distribution over the population. However, since APSRO, like PSRO, adds pure best responses, it may also require a large population to converge. 

In this work, we build on APSRO by aiming to add to the population in each iteration the myopically optimal policy, that is, the policy that maximally lowers the exploitability of the least-exploitable distribution of the resulting population. A key insight is that mixed strategies (i.e. stochastic policies) can lower the exploitability of a population more than pure strategies. To see this, note that a Nash equilibrium strategy is an optimal strategy to add, because the least-exploitable distribution over the resulting population will also be a Nash equilibrium strategy. If all Nash equilibria are mixed, as is often the case, then no pure strategy can be added to the population that reduces exploitability as much as the mixed strategy Nash equilibrium. 

Although finding the optimal strategy to add is as hard as solving the original game, we find that adding a rough approximation to the optimal strategy can offer striking empirical benefits in quickly reducing the exploitability of the restricted distribution. We present Self-Play PSRO (SP-PSRO), which similarly to APSRO learns a restricted distribution over the population via no regret against the opponent's best response. Additionally, SP-PSRO trains off-policy a new strategy against the opponent's best response. At the end of each iteration, SP-PSRO add two strategies to the population: (1) the time-average of the new strategy and (2) the best response to the opponent's restricted distribution. Section \ref{sec:sppsro} clarifies this algorithm using formal notation.

By training the new strategy off-policy, SP-PSRO learns both with no additional experience. Experiments on normal form games and extensive-form games such as Liar's Dice, Battleship, and Leduc Poker suggest that SP-PSRO can learn policies that are dramatically less exploitable than APSRO and PSRO.

\section{Background}
We consider extensive-form games with perfect recall \citep{hansen2004dynamic}. An extensive-form game progresses through a sequence of player actions, and has a \emph{world state} $w \in \mathcal{W}$ at each step. 
In an $N$-player game, $\mathcal{A} = \mathcal{A}_1 \times \cdots \times \mathcal{A}_N$ is the space of joint actions for the players. $\mathcal{A}_i(w) \subseteq \mathcal{A}_i$ denotes the set of legal actions for player $i \in \mathcal{N} = \{1, \ldots, N\}$ at world state $w$ and $a = (a_1, \ldots, a_N) \in \mathcal{A}$ denotes a joint action. At each world state, after the players choose a joint action, a transition function $\mathcal{T}(w, a) \in \Delta^\mathcal{W}$ determines the probability distribution of the next world state $w'$. Upon transition from world state $w$ to $w'$ via joint action $a$, player $i$ makes an \emph{observation} $o_i = \mathcal{O}_i(w,a,w')$. In each world state $w$, player $i$ receives a reward $\mathcal{R}_i(w)$. The game ends when the players reach a terminal world state. In this paper, we consider games that are guaranteed to end in a finite number of actions.

A \emph{history} is a sequence of actions and world states, denoted $h = (w^0, a^0, w^1, a^1, \ldots, w^t)$, where $w^0$ is the known initial world state of the game. $\mathcal{R}_i(h)$ and $\mathcal{A}_i(h)$ are, respectively, the reward and set of legal actions for player $i$ in the last world state of a history $h$. An \emph{information set} for player $i$, denoted by $s_i$, is a sequence of that player's observations and actions up until that time $s_i(h) = (a_i^0, o_i^1, a_i^1, \ldots, o_i^t)$. Define the set of all information sets for player $i$ to be $\mathcal{I}_i$. 
The set of histories that correspond to an information set $s_i$ is denoted $\mathcal{H}(s_i) = \{ h: s_i(h) = s_i \}$, and it is assumed that they all share the same set of legal actions $\mathcal{A}_i(s_i(h)) = \mathcal{A}_i(h)$. 

A player's \emph{strategy} $\pi_i$ 
is a function mapping from an information set to a probability distribution over actions. A \emph{strategy profile} $\pi$ is a tuple $(\pi_1, \ldots, \pi_N)$. All players other than $i$ are denoted $-i$, and their strategies are jointly denoted $\pi_{-i}$. A strategy for a history $h$ is denoted $\pi_i(h) = \pi_i(s_i(h))$ and $\pi(h)$ is the corresponding strategy profile. 
When a strategy $\pi_i$ is learned through RL, we refer to the learned strategy as a \emph{policy}.

The \emph{expected value (EV)} $v_i^{\pi}(h)$ for player $i$ is the expected sum of future rewards for player $i$ in history $h$, when all players play strategy profile $\pi$. The EV for an information set $s_i$ is denoted $v_i^{\pi}(s_i)$ and the EV for the entire game is denoted $v_i(\pi)$. A \emph{two-player zero-sum} game has $v_1(\pi) + v_2(\pi) = 0$ for all strategy profiles $\pi$. The EV for an action in an information set is denoted $v_i^{\pi}(s_i,a_i)$. A \emph{Nash equilibrium (NE)} is a strategy profile such that, if all players played their NE strategy, no player could achieve higher EV by deviating from it. Formally, $\pi^*$ is a NE if $v_i(\pi^*) = \max_{\pi_i}v_i(\pi_i, \pi^*_{-i})$ for each player $i$.

The \emph{exploitability} $e(\pi)$ of a strategy profile $\pi$ is defined as $e(\pi) = \sum_{i \in \mathcal{N}} \max_{\pi'_i}v_i(\pi'_i, \pi_{-i})$. A \emph{best response (BR)} strategy $\mathbb{BR}_i(\pi_{-i})$ for player $i$ to a strategy $\pi_{-i}$ is a strategy that maximally exploits $\pi_{-i}$: $\mathbb{BR}_i(\pi_{-i}) = \arg\max_{\pi_i}v_i(\pi_i, \pi_{-i})$. An \emph{$\bm{\epsilon}$-best response ($\bm{\epsilon}$-BR)} strategy $\mathbb{BR}^\epsilon_i(\pi_{-i})$ for player $i$ to a strategy $\pi_{-i}$ is a strategy that is at most $\epsilon$ worse for player $i$ than the best response: $v_i(\mathbb{BR}^\epsilon_i(\pi_{-i}), \pi_{-i}) \ge v_i(\mathbb{BR}_i(\pi_{-i}), \pi_{-i}) - \epsilon$. An \emph{$\bm{\epsilon}$-Nash equilibrium ($\bm{\epsilon}$-NE)} is a strategy profile $\pi$ in which, for each player $i$, $\pi_i$ is an $\epsilon$-BR to $\pi_{-i}$. 

A \emph{normal-form game} is a single-step extensive-form game. An extensive-form game induces a normal-form game in which the legal actions for player $i$ are its deterministic strategies $\bigtimes_{s_i \in \mathcal{I}_i} \mathcal{A}_i(s_i)$. These deterministic strategies are called \emph{pure strategies} of the normal-form game. A \emph{mixed strategy} is a distribution over a player's pure strategies. 

\section{Related Work}
Many recent works study the intersection of reinforcement learning and game theory. QPG~\citep{srinivasan2018actor} is an algorithm based on policy gradient that empirically converges to a NE when the learning rate is annealed. NeuRD~\citep{hennes2020neural}, Magnetic Mirror Descent~\citep{sokota2022unified}, and F-FoReL~\citep{perolat2021poincare} approximate replicator dynamics, mirror descent, and follow the regularized leader, respectively, with policy gradients. DeepNash, which is based on F-FoReL and NeuRD has achieved expert level performance at Stratego~\citep{perolat2022mastering}. Markov games generalize MDPs where players take simultaneous actions and observe the ground state of the game. Recent literature has shown that reinforcement learning algorithms converge to Nash equilibrium in two-player zero-sum Markov games~\citep{brafman2002r, wei2017online, perolat2018actor, xie2020learning, daskalakis2020independent, jin2021v} and in multi-player general-sum Markov potential games~\citep{leonardos2021global, mguni2021learning, fox2022independent, zhang2021gradient, ding2022independent}. Deep methods based on CFR~\citep{mcaleer2022escher, steinberger2020dream, deep_cfr} are another promising direction for scaling to large games. In this work we focus on a different set of deep RL algorithms for games based on PSRO. Advances made to PSRO can potentially be combined with the above methods via XDO~\citep{mcaleer2021xdo}. 

\subsection{Double Oracle (DO) and Policy Space Response Oracles (PSRO)}
Double Oracle~\citep{double_oracle} is an algorithm for finding a Nash equilibrium (NE) in normal-form games. The algorithm works by keeping a population of strategies $\Pi^t$ at time $t$. In each iteration, a NE $\pi^{*,t}$ is computed for the game restricted to strategies in $\Pi^t$. Then, a best response to this restricted NE for each player $\mathbb{BR}_i(\pi^{*,t}_{-i})$ is computed and added to the population $\Pi_i^{t+1} = \Pi_i^t \cup \{\mathbb{BR}_i(\pi^{*,t}_{-i}) \}$ for $i \in \{1, 2\}$. Although in the worst case DO must add all pure strategies, in many games DO empirically terminates early and outperforms alternative approaches. 

\begin{algorithm}[tb]
   \caption{Policy Space Response Oracle (PSRO) \cite{psro}}
   \label{psro}
\begin{algorithmic}
 \STATE {\bfseries Result:} Nash Equilibrium
   \STATE {\bfseries Input:} Initial population $\Pi^0$
   \REPEAT[for $t=0,1,\ldots$]
   \STATE $\pi^r \gets$ NE in game restricted to strategies in $\Pi^t$
  \FOR{$i \in \{1,2\}$}
     \FOR{$m$ iterations}
        \STATE Update policy $\beta_{-i}$ toward $\mathbb{BR}_{-i}(\pi^r_{i})$
     \ENDFOR
  \ENDFOR
  \STATE $\Pi^{t+1}_i \gets \Pi^t_i \cup \{\beta_i\}$ for $i \in \{1, 2\}$
  \UNTIL{No novel best response exists for either player}
   \STATE {\bfseries Return:} $\pi^r$
\end{algorithmic}
\end{algorithm}

Policy-Space Response Oracles (PSRO) \cite{psro} scales DO to large games by using reinforcement learning to approximate a best response. The restricted-game NE is computed on the empirical game matrix $U^\Pi$, generated by having each policy in the population $\Pi$ play each opponent policy and tracking average utility in a $\Pi_1 \times \Pi_2$ payoff matrix \citep{wellman2006methods}. PSRO is described in Algorithm \ref{psro}.

Several methods related to PSRO have been published in recent years. AlphaStar \citep{alphastar} trains a population of policies through a procedure that is somewhat similar to PSRO. AlphaStar also uses some elements of self-play when constructing its population, and outputs a population-restricted NE at test time. NXDO \citep{mcaleer2021xdo} iteratively adds reinforcement learning policies to a population but solves an extensive-form restricted game, which has been shown to be more efficient than solving a matrix-form restricted game as in PSRO. P2SRO \citep{mcaleer2020pipeline} parallelizes PSRO with convergence guarantees. Other work has looked at incorporating diversity \citep{liu2021towards, perez2021modelling} in the best response objective. However, since the best response is still pure, these methods suffer from the same problems of PSRO and APSRO as previously described. Other methods generalize PSRO to more players \citep{muller2019generalized, marris2021multi}, and meta-learn the restricted-NE population distribution \citep{feng2021neural}. \cite{slumbers2022learning} propose a PSRO-like approach for learning risk-averse equilibria.

\subsection{Anytime-PSRO (APSRO)}

\begin{algorithm}
\caption{Anytime PSRO (APSRO)  \citep{anytime_psro}}
 \label{anytime_psro_alg}
\begin{algorithmic}
 \STATE {\bfseries Result:} Approximate Nash Equilibrium
 \STATE {\bfseries Input:} Initial population $\Pi^0$
 \WHILE{Not Terminated \{$t=0,1,\ldots$\}}
 \STATE Initialize $\pi^r_i$ to uniform over $\Pi^t_i$ for $i \in \{1, 2\}$\;
 \FOR{$i \in \{1,2\}$} 
 \FOR{$n$ iterations}
 \FOR{$m$ iterations}
 \STATE Update policy $\beta_{-i}$ toward $\mathbb{BR}_{-i}(\pi^r_{i})$
 \ENDFOR
 \STATE Update $\pi^r_i$ via regret minimization vs. $\beta_{-i}$\;
 \ENDFOR
 \ENDFOR
 \STATE $\Pi^{t+1}_i = \Pi^t_i \cup \{\beta_i\}$ for $i \in \{1, 2\}$\;
 \ENDWHILE 
 \STATE {\bfseries Return:} $\pi^r$
\end{algorithmic}
\end{algorithm}

Although PSRO is guaranteed to converge to a NE by, at worst, enumerating all pure strategies, before convergence the exploitability of the restricted NE distribution may increase arbitrarily. This becomes a problem in large games where full enumeration is infeasible due to the large number of pure strategies. Anytime PSRO (APSRO) \citep{anytime_psro} is a PSRO-type algorithm that is guaranteed, up to approximation error, to not increase exploitability from one iteration to the next. 

In each iteration, APSRO creates for each player a different restricted game than PSRO does. The restricted game $G^i(\Pi)$ for player $i$ is created by restricting that player to only play strategies included in their population $\Pi_i$, while the opponent can play any strategy in the full game. The game value of $G^i(\Pi)$ for player $i$ is 

\begin{equation}
\label{restricted_game}
\max_{\pi_i \in \Pi_i}\min_{\pi_{-i}}v_i(\pi_i, \pi_{-i}).   
\end{equation}

The restricted game $G^i(\Pi)$ for player $i$ is then approximately solved by learning a restricted distribution $\pi_i^r$ over the population policies $\Pi_i$ via a no-regret algorithm against the opponent's best response $\beta_{-i}$, at the same time that the latter is trained with reinforcement learning. A description of APSRO is given in algorithm \ref{anytime_psro_alg}.

\section{Self-Play PSRO}\label{sec:sppsro}
Although APSRO does not increase exploitability from one iteration to the next by adding to each player's population the pure-strategy best response $\beta_i$ to the opponent's restricted distribution $\pi^r_{-i}$, it is not guaranteed to \emph{decrease} exploitability. $\beta_i$ may not be the myopically optimal pure strategy whose addition to $\Pi_i$ decreases exploitability the most. Moreover, adding mixed strategies can generally be better than adding pure strategies. 
\begin{figure}
    \centering
    \includegraphics[width=\textwidth]{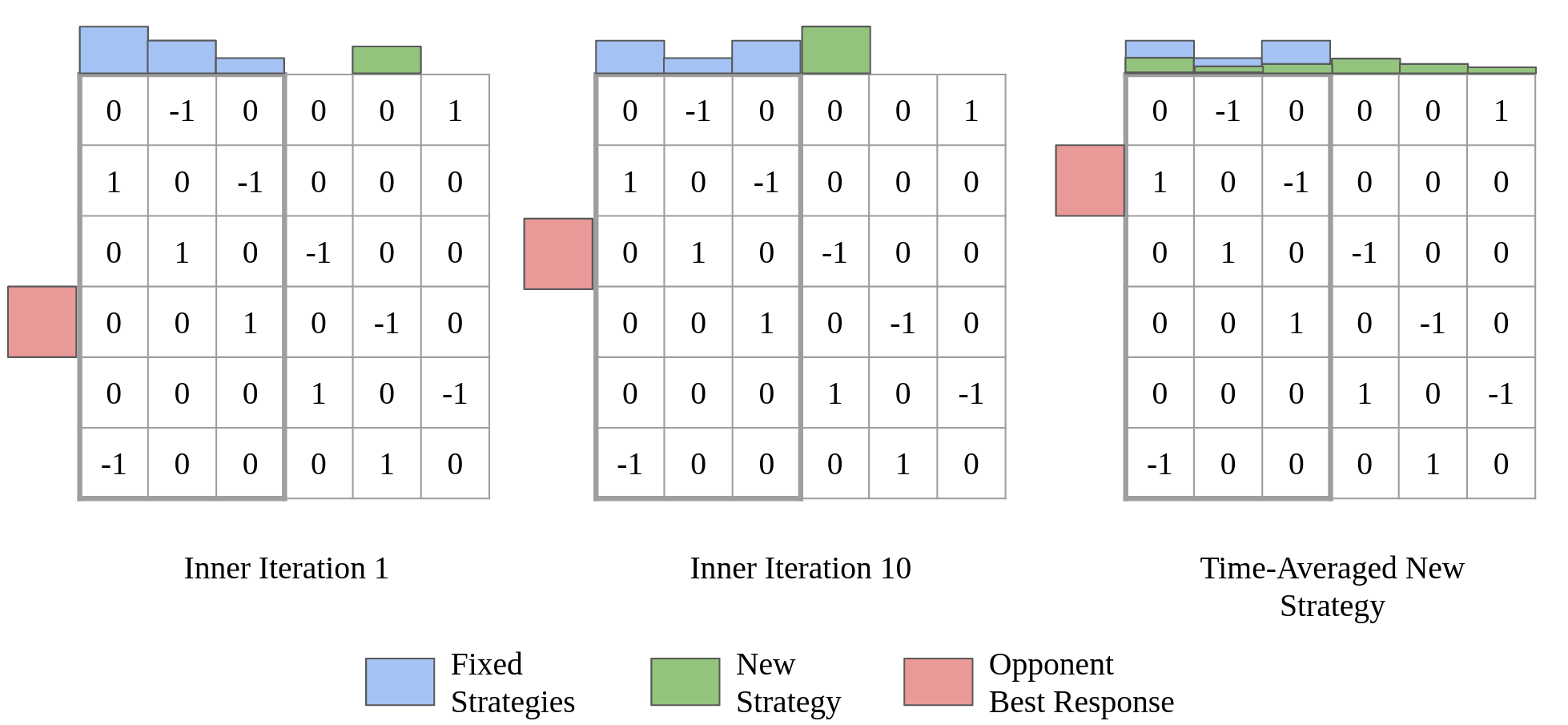}
    \caption{SP-PSRO. In this diagram we show how SP-PSRO works within an iteration from the perspective of the column player. The fixed population is shown in blue and the new strategy is shown in green. Every inner iteration, three things happen. (1) The opponent best response updates toward a best response to the current distribution over both the fixed population and the new strategy. (2) The new strategy updates toward a best response against the opponent best response. (3) The restricted distribution updates via no regret against the opponent best response. In the final iteration, the time-average of the new strategy and the player's best response to the opponent's restricted distribution (which is trained in a symmetric manner) are added to the population and the cycle starts again. }
    \label{fig:SP-PSRO}
\end{figure}

\begin{wrapfigure}{r}{0.33\textwidth}
  \begin{center}
    \includegraphics[width=0.3\textwidth]{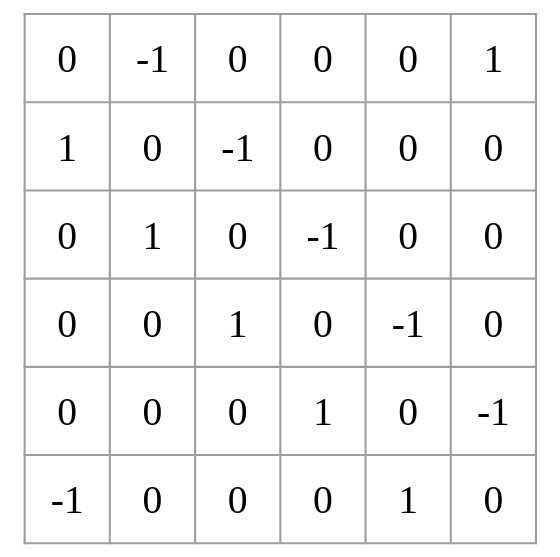} 
  \end{center}
    \caption{Big RPS Game. Any algorithm that only adds pure best responses, such as PSRO or APSRO, will expand all pure strategies before converging.}
    \label{fig:DO_Bad_Case}
\end{wrapfigure}

For example, consider the generalized Rock–Paper–Scissors game shown in Figure \ref{fig:DO_Bad_Case}. 
In this game, the NE mixes equally over all pure strategies. As a result, any DO method that only adds pure strategies will have to enumerate all pure strategies in the game before supporting the NE. 


Ideally, we would like to add a mixed strategy that decreases exploitability the most. A single-iteration objective would then be to find the strategy such that after it is added to the population and the least-exploitable distribution is computed over this new population, the exploitability of the resulting distribution is the lowest. In this example game, a mixed strategy that mixes over the pure strategies equally is optimal and will lower exploitability more than any pure strategy. The following proposition makes this point more explicit. 


Although the Nash equilibrium of the original game is the optimal mixed strategy to add to the population, finding a Nash equilibrium of the original game is very expensive and is our main goal in the first place. However, even if the mixed strategy we add is not a Nash equilibrium strategy, the fact that it is a mixed strategy instead of a pure strategy empirically decreases exploitability faster in early iterations than pure strategies. 
    
But we can potentially do better than just adding a random mixed strategy. By trying to roughly approximate a Nash equilibrium of the original game, we can expect to improve our population exploitability more than random. In other words, the closer the new strategy is to being a Nash equilibrium of the original game, the more it will lower the resulting exploitability of the population. 


Motivated by this, we propose Self-Play PSRO, a PSRO method that learns a new mixed strategy by best-responding to the opponent best response via off-policy reinforcement learning. To heuristically better approximate a Nash equilibrium, after the iteration has finished we output the time-average of the new strategy during self play. 

\begin{algorithm}
\caption{Self-Play PSRO}
\begin{algorithmic}
 \STATE {\bfseries Result:} Approximate Nash Equilibrium
 \STATE {\bfseries Input:} Initial population $\Pi^0 \cup \{\nu_i\}$
 \WHILE{Not Terminated \{$t=0,1,\ldots$\}}
 \STATE Initialize new strategy $\nu_{i}$ arbitrarily\;
 \STATE Initialize $\pi^r_i$ to uniform over $\Pi^t_i \cup\{\nu_i\}$ for $i \in \{1, 2\}$\;
 \FOR{$i \in \{1,2\}$} 
 \FOR{$n$ iterations}
 \FOR{$m$ iterations}
 \STATE Update policy $\beta_{-i}$ toward $\mathbb{BR}_{-i}(\pi^r_{i})$
 \STATE Update new strategy $\nu_{i}$ toward $\mathbb{BR}_{i}(\beta_{-i})$

 \ENDFOR
 \STATE Update $\pi^r_i$ via regret minimization vs. $\beta_{-i}$\;
 \ENDFOR
 \ENDFOR
 \STATE $\Pi^{t+1}_i = \Pi^t_i \cup \{\beta_i, \bar{\nu}_i\}$ for $i \in \{1, 2\}$\;
 \ENDWHILE 
 \STATE {\bfseries Return:} $\pi^r$
\end{algorithmic}
\end{algorithm}

SP-PSRO works by maintaining a restricted distribution $\pi_i^r$ over a population. Unlike PSRO, where $\pi_i^r$ is the NE of the restricted game, SP-PSRO trains $\pi_i^r$ as in APSRO, via regret minimization. In addition, at the beginning of each iteration, a new strategy $\nu_i$ is initialized and added to the population.

\begin{figure*}[t]
    \centering
    \begin{subfigure}[b]{0.32\textwidth}
        \centering
        \includegraphics[width=\textwidth]{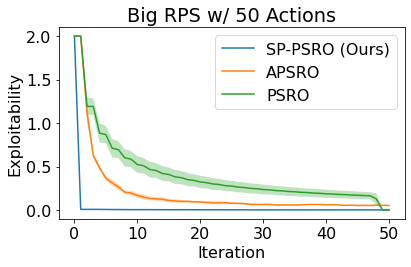}
        \caption{Big RPS with 50 Actions}
        \label{fig:big_rps}
    \end{subfigure}
    \begin{subfigure}[b]{0.32\textwidth}
        \centering
        \includegraphics[width=\textwidth]{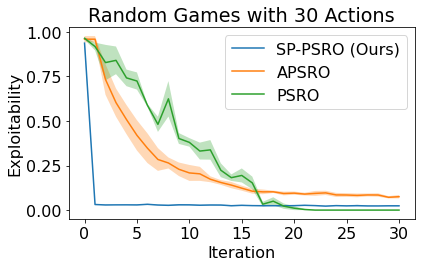}
        \caption{Random Games with 30 Actions}
        \label{fig:random_50}
    \end{subfigure}
    \begin{subfigure}[b]{0.32\textwidth}
        \centering
        \includegraphics[width=\textwidth]{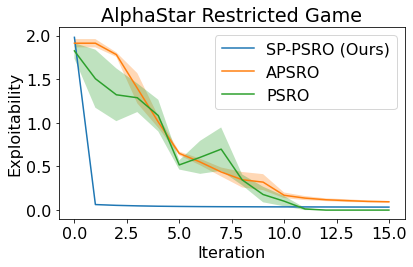}
        \caption{AlphaStar Restricted Game}
        \label{fig:alphastar_results}
    \end{subfigure}
     
     \medskip
    \centering
    \begin{subfigure}[b]{0.32\textwidth}
        \centering
        \includegraphics[width=\textwidth]{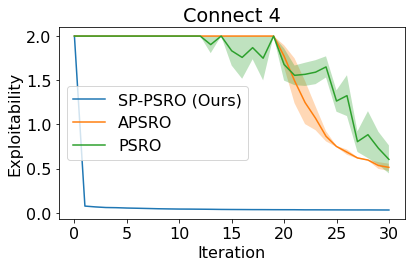}
        \caption{Connect 4 Restricted Game}
        \label{fig:connect_4}
    \end{subfigure}
    \begin{subfigure}[b]{0.32\textwidth}
        \centering
        \includegraphics[width=\textwidth]{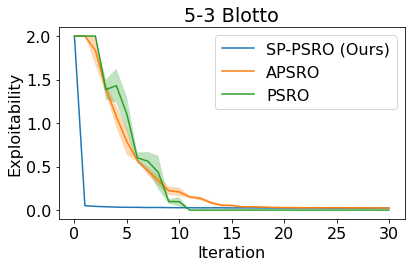}
        \caption{5-3 Blotto}
        \label{fig:blotto}
    \end{subfigure}
    \begin{subfigure}[b]{0.32\textwidth}
        \centering
        \includegraphics[width=\textwidth]{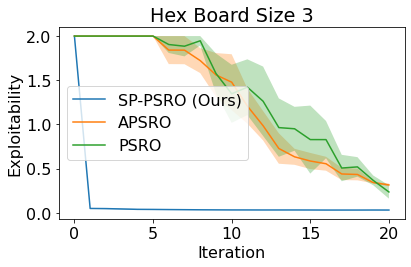}
        \caption{Hex-3 Restricted Game}
        \label{fig:hex}
    \end{subfigure}
     \caption{Normal-form games}
\end{figure*}

During an iteration, three training processes unfold concurrently. First, as in APSRO, the opponent's best response $\beta_i$ takes multiple update steps toward a best response to the current restricted distribution $\mathbb{BR}_{-i}(\pi^r_{i})$. Second, the new strategy $\nu_i$ is updated toward a best response to the opponent best response $\mathbb{BR}_{i}(\beta_{-i})$. Third, the restricted distribution $\pi_i^r$ is trained via regret minimization; this includes updating the probability of the new population strategy $\nu_i$, even as $\nu_i$ is also trained. This procedure can be thought of a form of self-play, in which the new strategy is updating against the opponent best response, while the opponent best response is updating against the restricted distribution, which also contains the new strategy. When the episode is terminated, the time-average $\bar{\nu_i}$ of $\nu_i$ is added to the population. Averaging over the updates of $\nu_i$ can be accomplished by checkpointing the policy over time and uniformly sampling checkpoints, or by training a neural network to distill a buffer of experience generated by $\nu_i$ as it trains. Since the new strategy is trained via off-policy reinforcement learning, SP-PSRO uses the exact same amount of environment experience as APSRO, but does require more compute to train the new network. 


\section{Experiments}
\subsection{Normal Form Experiments}
In this section we describe experiments on normal form games. To emulate the process of a strategy $\pi$ learning a best response to another policy $\pi'$, in every inner loop iteration $t$ we update $\pi$ by the following learning rule: $\pi_{t+1} = (1-\lambda)\pi_t + \lambda\times\mathbb{BR}_i(\pi')$. We consider six normal form games. The first, described in Figure \ref{fig:DO_Bad_Case}, is a large generalized Rock–Paper–Scissors game. The second is a random normal-form games generated by sampling each entry in the payoff matrix independently from a standard uniform distribution. The third game is the final restricted game of the AlphaStar population \citep{alphastar}. The next three are taken from \cite{perez2021modelling} and include a connect 4 restricted game, 5-3 blotto, a Hex restricted game.

\subsection{Tabular Experiments}

We evaluated SP-PSRO with tabular methods in a variety of games. We applied tabular SP-PSRO to the domains of Leduc Poker (9,457 states), a tiny version of Battleship (1,573 states), repeated Rock Paper Scissors (9,841 states), and a small version of Goofspiel (18,426 states). The experiments used game implementations and tools from the OpenSpiel library~\citep{lanctot2019openspiel}.

\begin{figure*}[t]
    \centering
    \begin{subfigure}[b]{0.4\textwidth}
        \centering
        \includegraphics[width=\textwidth]{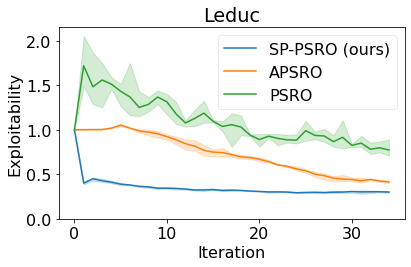}
        \caption{Leduc Poker}
        \label{fig:tabular_leduc}
    \end{subfigure}
    \begin{subfigure}[b]{0.4\textwidth}
        \centering
        \includegraphics[width=\textwidth]{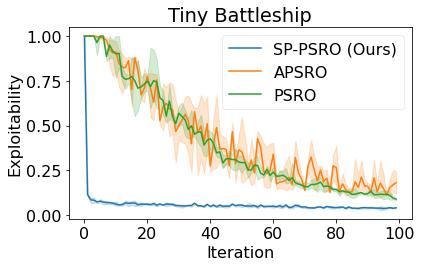}
        \caption{Battleship}
        \label{fig:tabular_battleship}
    \end{subfigure}
    \begin{subfigure}[b]{0.4\textwidth}
        \centering
        \includegraphics[width=\textwidth]{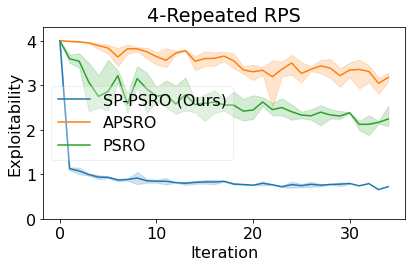}
        \caption{Repeated RPS}
        \label{fig:tabular_rps}
    \end{subfigure}
    \begin{subfigure}[b]{0.4\textwidth}
        \centering
        \includegraphics[width=\textwidth]{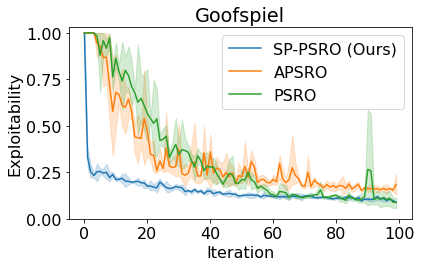}
        \caption{Goofspiel}
        \label{fig:tabular_goofspiel}
    \end{subfigure}
    \caption{Extensive-form games with tabular Q-learning best responses}
\end{figure*}

\subsubsection{Results}
SP-PSRO outperforms APSRO and PSRO in Leduc Poker (Figure \ref{fig:tabular_leduc}). We see a drastic improvement in performance starting in the first iteration. The hyperparameters for APSRO are the same as given in \cite{anytime_psro}. SP-PSRO outperforms both APSRO and PSRO in the small Battleship game (Figure \ref{fig:tabular_battleship}), 4-repeated Rock Paper Scissors (Figure~\ref{fig:tabular_rps}), and the small Goofspiel game (Figure \ref{fig:tabular_goofspiel}). In these 3 games, we use the same amount of compute for APSRO and SP-PSRO.

\subsubsection{Tabular Experiment Details}
In these experiments, the new population strategy $\nu_i$ and the BR $\beta_{-i}$ are the policies of tabular Q-learning agents. The tabular Q-learning agents are $\epsilon$-greedy. When training the Q-learning agent for $\beta_{-i}$, the episodes are also used to train the agent for $\nu_i$, in an off-policy manner. We use a constant value of $\epsilon$ for both agents. Note that this means that each agent learns to best-respond to the $\epsilon$-greedy agent of the other, instead of the underlying policy of the other. However, this approach uses half as many training episodes as would be needed if each trained independently, and still performs well.

\subsection{Deep Reinforcement Learning Experiments}
\begin{figure*}[t]
    \centering
    \begin{subfigure}[b]{0.32\textwidth}
        \centering
        \includegraphics[width=\textwidth]{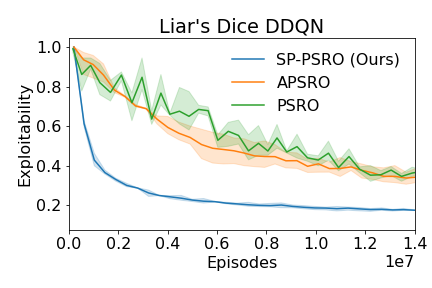}
        \caption{DRL Liars Dice}
        \label{fig:drl_liars_dice}
    \end{subfigure}
    \begin{subfigure}[b]{0.32\textwidth}
        \centering
        \includegraphics[width=\textwidth]{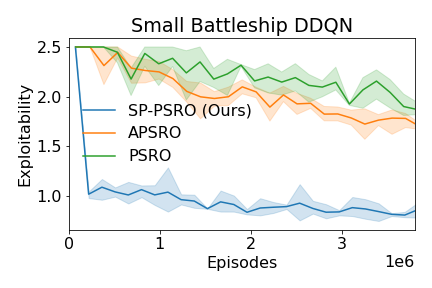}
        \caption{DRL Battleship}
        \label{fig:drl_battleship}
    \end{subfigure}
    \begin{subfigure}[b]{0.32\textwidth}
        \centering
        \includegraphics[width=\textwidth]{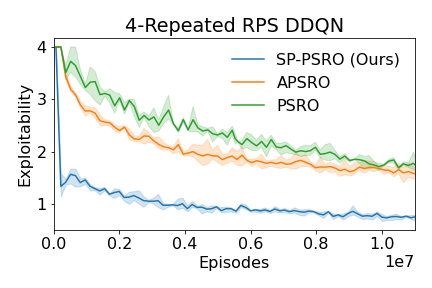}
        \caption{DRL Repeated RPS}
        \label{fig:drl_rps}
    \end{subfigure}
    \caption{    \label{fig:drl}
Extensive-form games with DDQN best responses}
\end{figure*}

When using deep reinforcement learning best-response operators with DDQN \cite{van2016deep}, SP-PSRO outperforms APSRO and PSRO in terms of sample efficiency (Figure \ref{fig:drl}). Tested on Liar's Dice, a small version of Battleship, and 4x Repeated RPS, SP-PSRO sees a significant improvement against other baselines in early-iteration exploitability. This early exploitability advantage seen by SP-PSRO is especially present in repeated RPS (Figure \ref{fig:drl_rps}), where the relative performance seen with deep RL methods roughly matches that of tabular methods.

\section{Discussion}
\subsection{Limitations}
One limitation of SP-PSRO is that if the new strategies happen to not be useful, including the new strategy can hurt the performance of the restricted distribution. This is primarily because it is harder to learn a no-regret distribution when one of the arms is changing, and secondly because including more actions (strategies) makes it harder for the no regret algorithm as well. 
A related limitation of SP-PSRO is that because including the new strategy makes it harder to learn the restricted distribution, we find that SP-PSRO tends to plateau higher than APSRO and can even slightly increase exploitability. To mitigate this, we switch over to APSRO after some iterations, but we have not introduced a principled method of determining when is a good time to switch.  
A third limitation of our method is that extra compute needs to be used to train the new strategy $\nu$. Also, if the average strategy is computed via supervised learning on a replay buffer of experience, this adds additional compute requirements to the algorithm. Lastly, SP-PSRO is somewhat unprincipled in that the new strategy generated by the self-play procedure is not guaranteed to be close to a Nash equilibrium, even in the time average. In the following section we speculate on interesting research directions to improve this part of SP-PSRO. 

\subsection{Future Work}
SP-PSRO opens up exciting connections to the literature regarding learning approximate Nash equilibria in large games. In particular, although we introduce an unprincipled self play method for approximating a Nash equilibrium, future work can find better ways of creating a new strategy that will better approximate a Nash equilibrium and therefore result in lower exploitability every iteration. For example, the data collected via the the opponent best response training against the restricted distribution can be used in a Monte-Carlo CFR-type algorithm to minimize regret on information sets visited during training. 

These directions open up the possibility of deriving the first regret bounds for double oracle algorithms that do not rely on the size of the effective pure strategy set \citep{dinh2021online}. It also introduces the possibility of combining the deep reinforcement learning from the best response with methods based on deep CFR. For example, perhaps the Q networks learned from the best responses can be used to minimize regret for the new strategy. 

Finally, our algorithm is a normal-form algorithm in that it mixes at the root of the game tree. \cite{mcaleer2021xdo} showed that this can be exponentially bad in the worst case, and introduced tabular (XDO) and deep (NXDO) algorithms to fix this problem. An interesting future direction is combining SP-PSRO with XDO and NXDO.

\bibliography{main.bib}

\begin{thebibliography}{41}
\providecommand{\natexlab}[1]{#1}
\providecommand{\url}[1]{\texttt{#1}}
\expandafter\ifx\csname urlstyle\endcsname\relax
  \providecommand{\doi}[1]{doi: #1}\else
  \providecommand{\doi}{doi: \begingroup \urlstyle{rm}\Url}\fi

\bibitem[Brafman \& Tennenholtz(2002)Brafman and Tennenholtz]{brafman2002r}
Brafman, R.~I. and Tennenholtz, M.
\newblock R-max-a general polynomial time algorithm for near-optimal
  reinforcement learning.
\newblock \emph{Journal of Machine Learning Research}, 3\penalty0
  (Oct):\penalty0 213--231, 2002.

\bibitem[Brown(1951)]{fp}
Brown, G.~W.
\newblock Iterative solution of games by fictitious play.
\newblock \emph{Activity analysis of production and allocation}, pp.\
  374--376, 1951.

\bibitem[Brown et~al.(2019)Brown, Lerer, Gross, and Sandholm]{deep_cfr}
Brown, N., Lerer, A., Gross, S., and Sandholm, T.
\newblock Deep counterfactual regret minimization.
\newblock In \emph{International Conference on Machine Learning}, pp.\
  793--802, 2019.

\bibitem[Daskalakis et~al.(2020)Daskalakis, Foster, and
  Golowich]{daskalakis2020independent}
Daskalakis, C., Foster, D.~J., and Golowich, N.
\newblock Independent policy gradient methods for competitive reinforcement
  learning.
\newblock \emph{Advances in neural information processing systems},
  33:\penalty0 5527--5540, 2020.

\bibitem[Ding et~al.(2022)Ding, Wei, Zhang, and
  Jovanovi{\'c}]{ding2022independent}
Ding, D., Wei, C.-Y., Zhang, K., and Jovanovi{\'c}, M.~R.
\newblock Independent policy gradient for large-scale markov potential games:
  Sharper rates, function approximation, and game-agnostic convergence.
\newblock \emph{arXiv preprint arXiv:2202.04129}, 2022.

\bibitem[Dinh et~al.(2021)Dinh, Yang, Tian, Nieves, Slumbers, Mguni, Ammar, and
  Wang]{dinh2021online}
Dinh, L.~C., Yang, Y., Tian, Z., Nieves, N.~P., Slumbers, O., Mguni, D.~H.,
  Ammar, H.~B., and Wang, J.
\newblock Online double oracle.
\newblock \emph{arXiv preprint arXiv:2103.07780}, 2021.

\bibitem[Feng et~al.(2021)Feng, Slumbers, Wan, Liu, McAleer, Wen, Wang, and
  Yang]{feng2021neural}
Feng, X., Slumbers, O., Wan, Z., Liu, B., McAleer, S., Wen, Y., Wang, J., and
  Yang, Y.
\newblock Neural auto-curricula in two-player zero-sum games.
\newblock \emph{Advances in Neural Information Processing Systems}, 34, 2021.

\bibitem[Fox et~al.(2022)Fox, Mcaleer, Overman, and
  Panageas]{fox2022independent}
Fox, R., Mcaleer, S.~M., Overman, W., and Panageas, I.
\newblock Independent natural policy gradient always converges in markov
  potential games.
\newblock In \emph{International Conference on Artificial Intelligence and
  Statistics}, pp.\  4414--4425. PMLR, 2022.

\bibitem[Hansen et~al.(2004)Hansen, Bernstein, and
  Zilberstein]{hansen2004dynamic}
Hansen, E.~A., Bernstein, D.~S., and Zilberstein, S.
\newblock Dynamic programming for partially observable stochastic games.
\newblock \emph{Conference on Artificial Intelligence (AAAI)}, 2004.

\bibitem[Heinrich \& Silver(2016)Heinrich and Silver]{nfsp}
Heinrich, J. and Silver, D.
\newblock Deep reinforcement learning from self-play in imperfect-information
  games.
\newblock \emph{arXiv preprint arXiv:1603.01121}, 2016.

\bibitem[Hennes et~al.(2020)Hennes, Morrill, Omidshafiei, Munos, Perolat,
  Lanctot, Gruslys, Lespiau, Parmas, Du{\'e}{\~n}ez-Guzm{\'a}n,
  et~al.]{hennes2020neural}
Hennes, D., Morrill, D., Omidshafiei, S., Munos, R., Perolat, J., Lanctot, M.,
  Gruslys, A., Lespiau, J.-B., Parmas, P., Du{\'e}{\~n}ez-Guzm{\'a}n, E.,
  et~al.
\newblock Neural replicator dynamics: Multiagent learning via hedging policy
  gradients.
\newblock In \emph{Proceedings of the 19th International Conference on
  Autonomous Agents and MultiAgent Systems}, pp.\  492--501, 2020.

\bibitem[Jin et~al.(2021)Jin, Liu, Wang, and Yu]{jin2021v}
Jin, C., Liu, Q., Wang, Y., and Yu, T.
\newblock V-learning--a simple, efficient, decentralized algorithm for
  multiagent rl.
\newblock \emph{arXiv preprint arXiv:2110.14555}, 2021.

\bibitem[Kingma \& Ba(2014)Kingma and Ba]{kingma2014adam}
Kingma, D.~P. and Ba, J.
\newblock Adam: A method for stochastic optimization.
\newblock \emph{arXiv preprint arXiv:1412.6980}, 2014.

\bibitem[Lanctot et~al.(2017)Lanctot, Zambaldi, Gruslys, Lazaridou, Tuyls,
  P{\'e}rolat, Silver, and Graepel]{psro}
Lanctot, M., Zambaldi, V., Gruslys, A., Lazaridou, A., Tuyls, K., P{\'e}rolat,
  J., Silver, D., and Graepel, T.
\newblock A unified game-theoretic approach to multiagent reinforcement
  learning.
\newblock In \emph{Advances in Neural Information Processing Systems
  (NeurIPS)}, 2017.

\bibitem[Lanctot et~al.(2019)Lanctot, Lockhart, Lespiau, Zambaldi, Upadhyay,
  P{\'e}rolat, Srinivasan, Timbers, Tuyls, Omidshafiei,
  et~al.]{lanctot2019openspiel}
Lanctot, M., Lockhart, E., Lespiau, J.-B., Zambaldi, V., Upadhyay, S.,
  P{\'e}rolat, J., Srinivasan, S., Timbers, F., Tuyls, K., Omidshafiei, S.,
  et~al.
\newblock Openspiel: A framework for reinforcement learning in games.
\newblock \emph{arXiv preprint arXiv:1908.09453}, 2019.

\bibitem[Leonardos et~al.(2021)Leonardos, Overman, Panageas, and
  Piliouras]{leonardos2021global}
Leonardos, S., Overman, W., Panageas, I., and Piliouras, G.
\newblock Global convergence of multi-agent policy gradient in markov potential
  games.
\newblock \emph{arXiv preprint arXiv:2106.01969}, 2021.

\bibitem[Liang et~al.(2018)Liang, Liaw, Nishihara, Moritz, Fox, Goldberg,
  Gonzalez, Jordan, and Stoica]{liang2018rllib}
Liang, E., Liaw, R., Nishihara, R., Moritz, P., Fox, R., Goldberg, K.,
  Gonzalez, J., Jordan, M., and Stoica, I.
\newblock Rllib: Abstractions for distributed reinforcement learning.
\newblock In \emph{International Conference on Machine Learning}, pp.\
  3053--3062. PMLR, 2018.

\bibitem[Liu et~al.(2021)Liu, Jia, Wen, Hu, Chen, Fan, Hu, and
  Yang]{liu2021towards}
Liu, X., Jia, H., Wen, Y., Hu, Y., Chen, Y., Fan, C., Hu, Z., and Yang, Y.
\newblock Towards unifying behavioral and response diversity for open-ended
  learning in zero-sum games.
\newblock \emph{Advances in Neural Information Processing Systems},
  34:\penalty0 941--952, 2021.

\bibitem[Marris et~al.(2021)Marris, Muller, Lanctot, Tuyls, and
  Graepel]{marris2021multi}
Marris, L., Muller, P., Lanctot, M., Tuyls, K., and Graepel, T.
\newblock Multi-agent training beyond zero-sum with correlated equilibrium
  meta-solvers.
\newblock In \emph{International Conference on Machine Learning}, pp.\
  7480--7491. PMLR, 2021.

\bibitem[McAleer et~al.(2020)McAleer, Lanier, Fox, and
  Baldi]{mcaleer2020pipeline}
McAleer, S., Lanier, J., Fox, R., and Baldi, P.
\newblock Pipeline {PSRO}: A scalable approach for finding approximate {Nash}
  equilibria in large games.
\newblock In \emph{Advances in Neural Information Processing Systems}, 2020.

\bibitem[McAleer et~al.(2021)McAleer, Lanier, Wang, Baldi, and
  Fox]{mcaleer2021xdo}
McAleer, S., Lanier, J.~B., Wang, K.~A., Baldi, P., and Fox, R.
\newblock {XDO}: A double oracle algorithm for extensive-form games.
\newblock \emph{Advances in Neural Information Processing Systems (NeurIPS)},
  2021.

\bibitem[McAleer et~al.(2022{\natexlab{a}})McAleer, Farina, Lanctot, and
  Sandholm]{mcaleer2022escher}
McAleer, S., Farina, G., Lanctot, M., and Sandholm, T.
\newblock Escher: Eschewing importance sampling in games by computing a history
  value function to estimate regret.
\newblock \emph{arXiv preprint arXiv:2206.04122}, 2022{\natexlab{a}}.

\bibitem[McAleer et~al.(2022{\natexlab{b}})McAleer, Wang, Lanier, Lanctot,
  Baldi, Sandholm, and Fox]{anytime_psro}
McAleer, S., Wang, K., Lanier, J.~B., Lanctot, M., Baldi, P., Sandholm, T., and
  Fox, R.
\newblock Anytime {PSRO} for two-player zero-sum games.
\newblock \emph{CoRR}, abs/2201.07700, 2022{\natexlab{b}}.
\newblock URL \url{https://arxiv.org/abs/2201.07700}.

\bibitem[McMahan et~al.(2003)McMahan, Gordon, and Blum]{double_oracle}
McMahan, H.~B., Gordon, G.~J., and Blum, A.
\newblock Planning in the presence of cost functions controlled by an
  adversary.
\newblock \emph{Proceedings of the 20th International Conference on Machine
  Learning (ICML)}, 2003.

\bibitem[Mguni et~al.(2021)Mguni, Wu, Du, Yang, Wang, Li, Wen, Jennings, and
  Wang]{mguni2021learning}
Mguni, D.~H., Wu, Y., Du, Y., Yang, Y., Wang, Z., Li, M., Wen, Y., Jennings,
  J., and Wang, J.
\newblock Learning in nonzero-sum stochastic games with potentials.
\newblock In \emph{International Conference on Machine Learning}, pp.\
  7688--7699. PMLR, 2021.

\bibitem[Muller et~al.(2020)Muller, Omidshafiei, Rowland, Tuyls, Perolat, Liu,
  Hennes, Marris, Lanctot, Hughes, et~al.]{muller2019generalized}
Muller, P., Omidshafiei, S., Rowland, M., Tuyls, K., Perolat, J., Liu, S.,
  Hennes, D., Marris, L., Lanctot, M., Hughes, E., et~al.
\newblock A generalized training approach for multiagent learning.
\newblock \emph{International Conference on Learning Representations (ICLR)},
  2020.

\bibitem[Perez-Nieves et~al.(2021)Perez-Nieves, Yang, Slumbers, Mguni, Wen, and
  Wang]{perez2021modelling}
Perez-Nieves, N., Yang, Y., Slumbers, O., Mguni, D.~H., Wen, Y., and Wang, J.
\newblock Modelling behavioural diversity for learning in open-ended games.
\newblock In \emph{International Conference on Machine Learning}, pp.\
  8514--8524. PMLR, 2021.

\bibitem[Perolat et~al.(2018)Perolat, Piot, and Pietquin]{perolat2018actor}
Perolat, J., Piot, B., and Pietquin, O.
\newblock Actor-critic fictitious play in simultaneous move multistage games.
\newblock In \emph{International Conference on Artificial Intelligence and
  Statistics}, pp.\  919--928. PMLR, 2018.

\bibitem[Perolat et~al.(2021)Perolat, Munos, Lespiau, Omidshafiei, Rowland,
  Ortega, Burch, Anthony, Balduzzi, De~Vylder, et~al.]{perolat2021poincare}
Perolat, J., Munos, R., Lespiau, J.-B., Omidshafiei, S., Rowland, M., Ortega,
  P., Burch, N., Anthony, T., Balduzzi, D., De~Vylder, B., et~al.
\newblock From {P}oincar{\'e} recurrence to convergence in imperfect
  information games: Finding equilibrium via regularization.
\newblock In \emph{International Conference on Machine Learning}, pp.\
  8525--8535. PMLR, 2021.

\bibitem[Perolat et~al.(2022)Perolat, de~Vylder, Hennes, Tarassov, Strub,
  de~Boer, Muller, Connor, Burch, Anthony, et~al.]{perolat2022mastering}
Perolat, J., de~Vylder, B., Hennes, D., Tarassov, E., Strub, F., de~Boer, V.,
  Muller, P., Connor, J.~T., Burch, N., Anthony, T., et~al.
\newblock Mastering the game of stratego with model-free multiagent
  reinforcement learning.
\newblock \emph{arXiv e-prints}, pp.\  arXiv--2206, 2022.

\bibitem[Slumbers et~al.(2022)Slumbers, Mguni, McAleer, Wang, and
  Yang]{slumbers2022learning}
Slumbers, O., Mguni, D.~H., McAleer, S., Wang, J., and Yang, Y.
\newblock Learning risk-averse equilibria in multi-agent systems.
\newblock \emph{arXiv preprint arXiv:2205.15434}, 2022.

\bibitem[Sokota et~al.(2022)Sokota, D'Orazio, Kolter, Loizou, Lanctot,
  Mitliagkas, Brown, and Kroer]{sokota2022unified}
Sokota, S., D'Orazio, R., Kolter, J.~Z., Loizou, N., Lanctot, M., Mitliagkas,
  I., Brown, N., and Kroer, C.
\newblock A unified approach to reinforcement learning, quantal response
  equilibria, and two-player zero-sum games.
\newblock \emph{arXiv preprint arXiv:2206.05825}, 2022.

\bibitem[Srinivasan et~al.(2018)Srinivasan, Lanctot, Zambaldi, P{\'e}rolat,
  Tuyls, Munos, and Bowling]{srinivasan2018actor}
Srinivasan, S., Lanctot, M., Zambaldi, V., P{\'e}rolat, J., Tuyls, K., Munos,
  R., and Bowling, M.
\newblock Actor-critic policy optimization in partially observable multiagent
  environments.
\newblock \emph{Advances in neural information processing systems}, 31, 2018.

\bibitem[Steinberger et~al.(2020)Steinberger, Lerer, and
  Brown]{steinberger2020dream}
Steinberger, E., Lerer, A., and Brown, N.
\newblock Dream: Deep regret minimization with advantage baselines and
  model-free learning.
\newblock \emph{arXiv preprint arXiv:2006.10410}, 2020.

\bibitem[Van~Hasselt et~al.(2016)Van~Hasselt, Guez, and Silver]{van2016deep}
Van~Hasselt, H., Guez, A., and Silver, D.
\newblock Deep reinforcement learning with double q-learning.
\newblock In \emph{AAAI conference on artificial intelligence}, volume~30,
  2016.

\bibitem[Vinyals et~al.(2019)Vinyals, Babuschkin, Czarnecki, Mathieu, Dudzik,
  Chung, Choi, Powell, Ewalds, Georgiev, et~al.]{alphastar}
Vinyals, O., Babuschkin, I., Czarnecki, W.~M., Mathieu, M., Dudzik, A., Chung,
  J., Choi, D.~H., Powell, R., Ewalds, T., Georgiev, P., et~al.
\newblock Grandmaster level in {StarCraft II} using multi-agent reinforcement
  learning.
\newblock \emph{Nature}, 575\penalty0 (7782):\penalty0 350--354, 2019.

\bibitem[Vitter(1985)]{vitter1985random}
Vitter, J.~S.
\newblock Random sampling with a reservoir.
\newblock \emph{ACM Transactions on Mathematical Software (TOMS)}, 11\penalty0
  (1):\penalty0 37--57, 1985.

\bibitem[Wei et~al.(2017)Wei, Hong, and Lu]{wei2017online}
Wei, C.-Y., Hong, Y.-T., and Lu, C.-J.
\newblock Online reinforcement learning in stochastic games.
\newblock \emph{Advances in Neural Information Processing Systems}, 30, 2017.

\bibitem[Wellman(2006)]{wellman2006methods}
Wellman, M.~P.
\newblock Methods for empirical game-theoretic analysis.
\newblock \emph{AAAI conference on artificial intelligence}, 2006.

\bibitem[Xie et~al.(2020)Xie, Chen, Wang, and Yang]{xie2020learning}
Xie, Q., Chen, Y., Wang, Z., and Yang, Z.
\newblock Learning zero-sum simultaneous-move markov games using function
  approximation and correlated equilibrium.
\newblock In \emph{Conference on learning theory}, pp.\  3674--3682. PMLR,
  2020.

\bibitem[Zhang et~al.(2021)Zhang, Ren, and Li]{zhang2021gradient}
Zhang, R., Ren, Z., and Li, N.
\newblock Gradient play in stochastic games: stationary points, convergence,
  and sample complexity.
\newblock \emph{arXiv preprint arXiv:2106.00198}, 2021.

\end{thebibliography}
\bibliographystyle{icml2021}

\appendix

\section{Additional Normal-Form Experiments}
\begin{figure*}[t]
    \centering
    \begin{subfigure}[b]{0.32\textwidth}
        \centering
        \includegraphics[width=\textwidth]{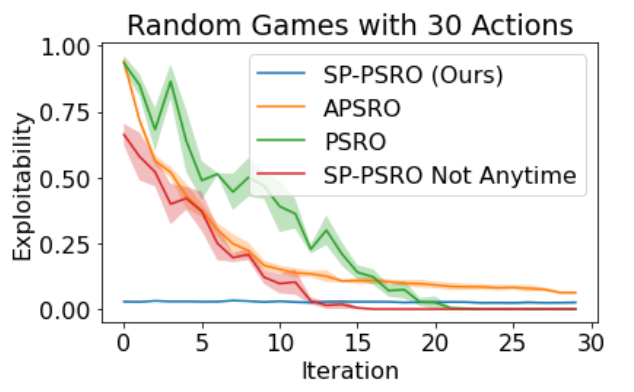}
        \caption{Random Games with 30 Actions}
        \label{fig:rg}
    \end{subfigure}
    \begin{subfigure}[b]{0.32\textwidth}
        \centering
        \includegraphics[width=\textwidth]{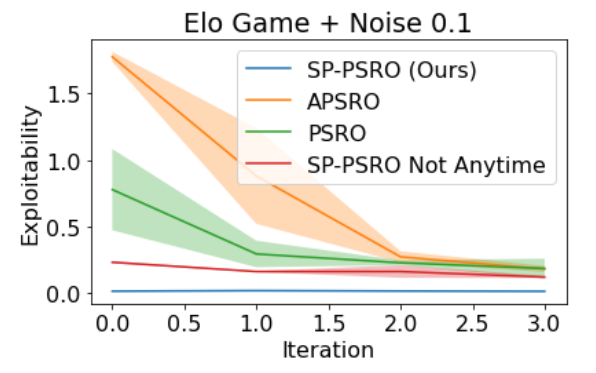}
        \caption{Elo Game + Noise 0.1}
        \label{fig:elo0.1}
    \end{subfigure}
    \begin{subfigure}[b]{0.32\textwidth}
        \centering
        \includegraphics[width=\textwidth]{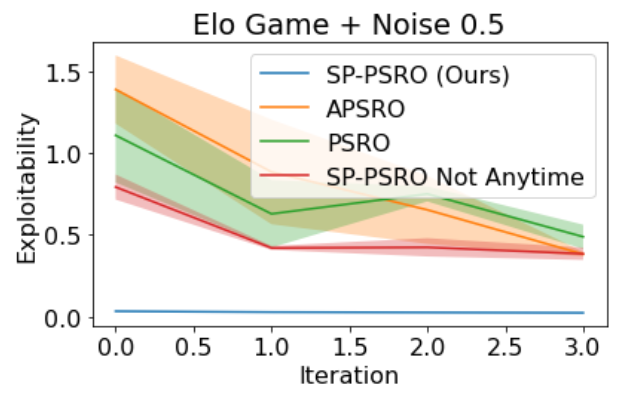}
        \caption{Elo Game + Noise 0.5}
        \label{fig:elo0.5}
    \end{subfigure}
    \begin{subfigure}[b]{0.32\textwidth}
        \centering
        \includegraphics[width=\textwidth]{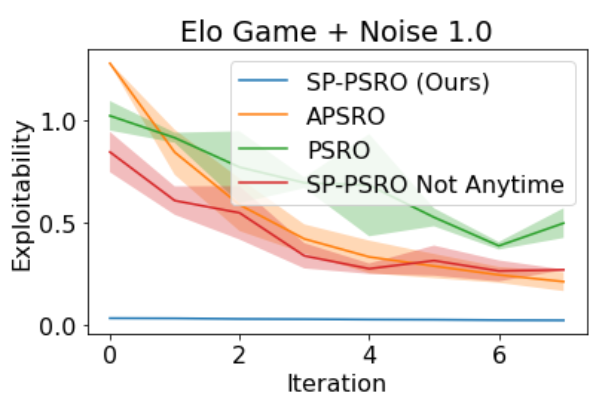}
        \caption{Elo Game + Noise 1.0}
        \label{fig:elo}
    \end{subfigure}
        \begin{subfigure}[b]{0.32\textwidth}
        \centering
        \includegraphics[width=\textwidth]{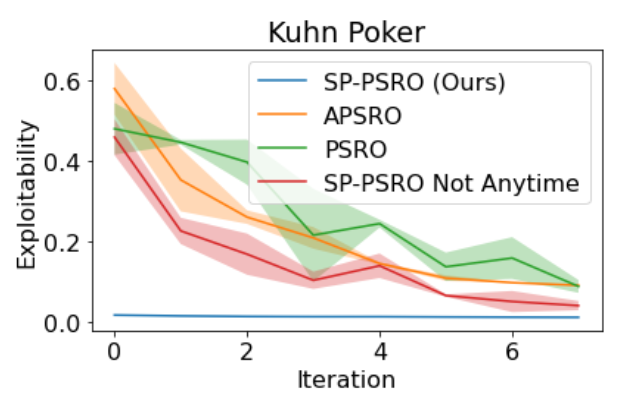}
        \caption{Kuhn Poker}
        \label{fig:kuhn_nf}
    \end{subfigure}
    \begin{subfigure}[b]{0.32\textwidth}
        \centering
        \includegraphics[width=\textwidth]{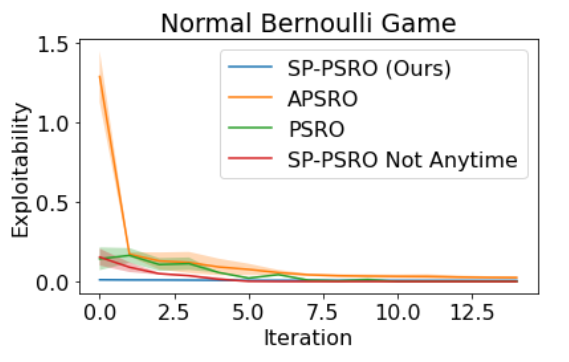}
        \caption{Normal Bernoulli Game}
        \label{fig:bernouilli}
    \end{subfigure}
    \begin{subfigure}[b]{0.32\textwidth}
        \centering
        \includegraphics[width=\textwidth]{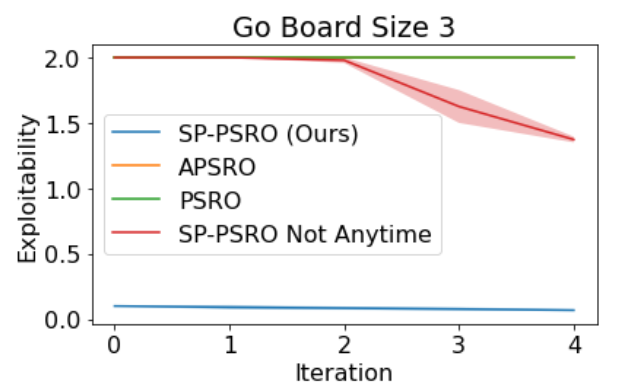}
        \caption{Go Board Size 3}
        \label{fig:go}
    \end{subfigure}
    \begin{subfigure}[b]{0.32\textwidth}
        \centering
        \includegraphics[width=\textwidth]{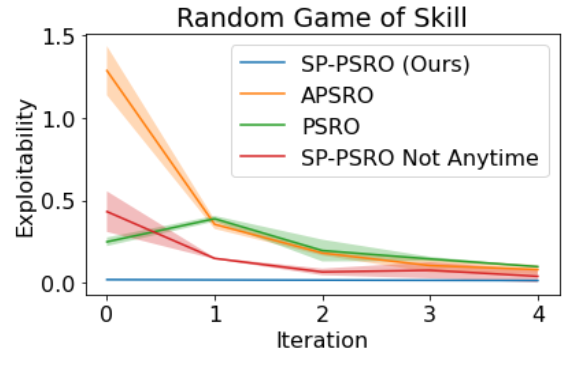}
        \caption{Random Game of Skill}
        \label{fig:rgos}
    \end{subfigure}
        \begin{subfigure}[b]{0.32\textwidth}
        \centering
        \includegraphics[width=\textwidth]{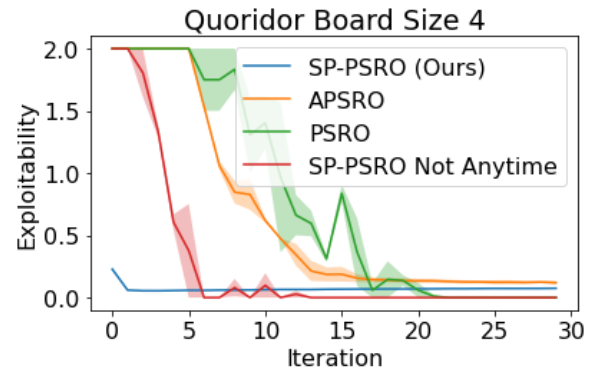}
        \caption{Quoridor Board Size 4}
        \label{fig:quoridor}
    \end{subfigure}
    \begin{subfigure}[b]{0.32\textwidth}
        \centering
        \includegraphics[width=\textwidth]{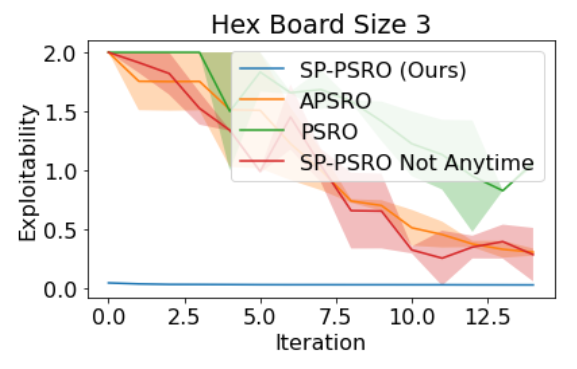}
        \caption{Hex Board Size 3}
        \label{fig:hex3}
    \end{subfigure}
    \begin{subfigure}[b]{0.32\textwidth}
        \centering
        \includegraphics[width=\textwidth]{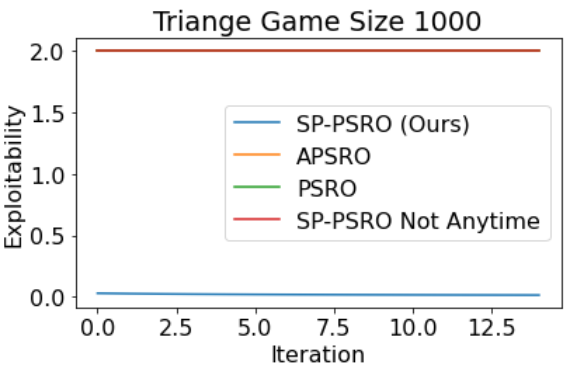}
        \caption{Triangle Game Size 1000}
        \label{fig:triangle1000}
    \end{subfigure}
    \begin{subfigure}[b]{0.32\textwidth}
        \centering
        \includegraphics[width=\textwidth]{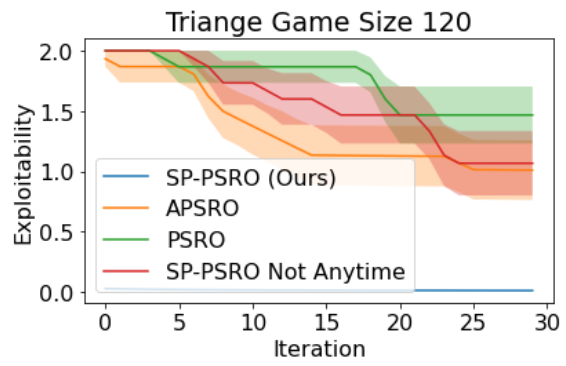}
        \caption{Triangle Game Size 120}
        \label{fig:triangle120}
    \end{subfigure}
    
    \caption{Additional Normal-Form Game Experiments}
\end{figure*}

In this section we report additional normal form game experiments. All games in this section are from \cite{perez2021modelling}. Note that the zeroth iteration is not included in the plots. Similar to the main results in the paper we find that SP-PSRO achieves much lower exploitability than existing PSRO based methods and does so much faster, across all games studied. We include an ablation, labeled SP-PSRO Not Anytime, that is the same as SP-PSRO in that it trains a new strategy to be a best response to the opponent best response, but unlike SP-PSRO does not update the restricted distribution via no-regret as in Anytime PSRO. As shown in the figures, Anytime PSRO is a crucial piece of SP-PSRO, and excluding this aspect results in much worse performance. We find that when Anytime PSRO is excluded, the opponent best response will be best responding to a static opponent, and the best response to this best response will tend to be a pure strategy. As a result, we do not get to explore the strategy space, and the average new strategy will simply be another pure strategy. In some games we see that SP-PSRO Not Anytime and PSRO converge to lower exploitability than SP-PSRO and Anytime PSRO. This is because SP-PSRO Not Anytime and PSRO both use exact meta-solvers, which return the exact Nash equilibrium upon convergence, while SP-PSRO and APSRO use a no-regret procedure to find the least-exploitable restricted distribution. 



\section{Additional Neural Experiments}
\begin{figure*}[t]
    \centering
    \begin{subfigure}[b]{0.32\textwidth}
        \centering
        \includegraphics[width=\textwidth]{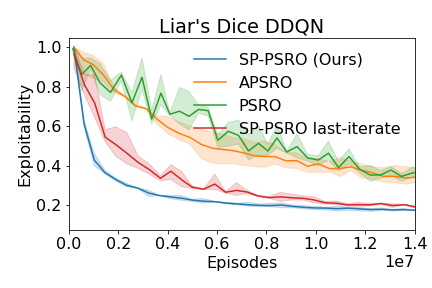}
        \caption{DRL Liars Dice}
        \label{fig:drl_liars_dice_li}
    \end{subfigure}
    \begin{subfigure}[b]{0.32\textwidth}
        \centering
        \includegraphics[width=\textwidth]{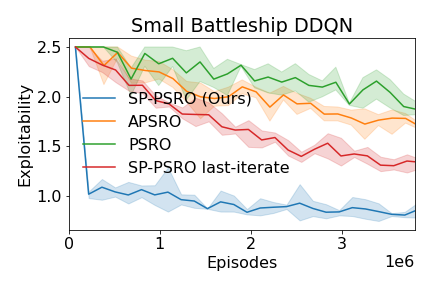}
        \caption{DRL Battleship}
        \label{fig:drl_battleship_li}
    \end{subfigure}
    \begin{subfigure}[b]{0.32\textwidth}
        \centering
        \includegraphics[width=\textwidth]{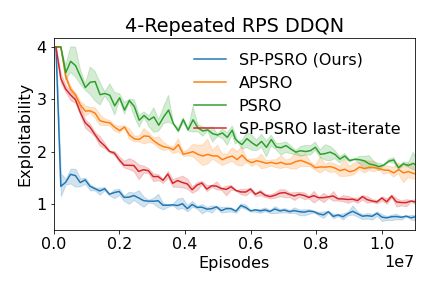}
        \caption{DRL Repeated RPS}
        \label{fig:drl_rps_li}
    \end{subfigure}
    \caption{\label{fig:drl_last_iterate}SP-PSRO last-iterate on extensive-form games with DDQN best responses}
\end{figure*}

We compare an alternate last-iterate version of SP-PSRO against the default SP-PSRO method and other baselines in Figure \ref{fig:drl_last_iterate}. In the SP-PSRO last-iterate variant, we add the pure-strategy weights of $\nu_i$ in its final RL iteration to the population rather than calculating and adding the time average $\bar{\nu_i}$. Exploitability is also calculated using $\nu_i$ rather than $\bar{\nu_i}$. SP-PSRO last-iterate improves upon APSRO and PSRO due to the additional, potentially useful, population policy. However, $\nu_i$ is less able to roughly approximate a NE because it represents a single pure-strategy approximate best-response to $\beta_{-i}$ rather than a mixture of multiple best-responses. Because of this, we still see an additional gain in exploitability versus sample-efficiency when transitioning from SP-PSRO last-iterate to the default time-average version of SP-PSRO.

\section{Extensive-Form Game Environments}

All extensive-form games tested with are from the OpenSpiel framework \citep{lanctot2019openspiel}, and can be loaded using OpenSpiel with the following parameters:

\begin{description}
\item[Leduc] Game Name: \verb|leduc_poker| \\
Parameters: \verb|{"players": 2}|
\item[Goofspiel] Game Name: \verb|goofspiel| \\
Parameters: \verb|{"imp_info": True, "num_cards": 5,|\\
            \verb|"points_order":"descending",}|
\item[Tiny Battleship] Game Name: \verb|battleship| \\
Parameters: \verb|{"board_width": 2, "board_height":  2,|\\
            \verb|"ship_sizes":  '[1]', "ship_values": '[1]',|\\
            \verb|"num_shots": 2, "allow_repeated_shots": False,}|
\item[Small Battleship] Game Name: \verb|battleship| \\
Parameters: \verb|{"board_width": 2, "board_height": 2, | \\
\verb|"ship_sizes": "[1;2]", "ship_values": "[1;2]",| \\
\verb|"num_shots": 4, "allow_repeated_shots": False}|

\item[4x Repeated RPS] Game Name: \verb|repeated_game| \\
Parameters: \verb|{"num_repetitions": 4, "enable_infostate": True, | \\
\verb|"stage_game": "matrix_rps"}|
\item[Liar's Dice] Game Name: \verb|liars_dice| \\
Parameters: \verb|None|
\end{description}

Goofspiel and Repeated RPS are converted from simultaneous-move games into turn-based games using OpenSpiel's \verb|convert_to_turn_based()| game transform. Repeated RPS is created from the \verb|matrix_rps| game using the \verb|create_repeated_game| game transform.

\section{Tabular Training Details}

In all tabular experiments, when calculating the payoff for a strategy profile ($v_i(\pi)$), the payoff is calculated exactly using a full tree traversal.

\subsection{SP-PSRO}
We performed a hyperparameter sweep to find the number of episodes per iteration and number of Exp3 updates per iteration which minimize exploitability in Leduc poker after 35 iterations of SP-PSRO (Table \ref{table:tabular-leduc}). We used these hyperparameters for SP-PSRO for tabular experiments in all games.

In each iteration, we split the Q-learning training and Exp3 updates into 600 equally-sized batches each, and alternate between a batch of Exp3 updates and a batch of Q-learning episodes until the end of the iteration. For example, when using $799,800$ episodes and $19,800$ Exp3 updates per iteration, we repeat the following 600 times: for each player, perform $799,800/600 = 1333$ Q-learning episodes and then $19,800/600 = 33$ Exp3 updates.

\begin{table}[H]
\centering
\begin{tabular}{ll}
episodes per iteration & 799,800\\
Exp3 updates per iteration & 19,800\\
Q-learning learning rate & 0.025\\
Q-learning exploration $\epsilon$ & Constant, 0.2
\end{tabular}
\caption{Tabular experiment details}
\label{table:tabular-leduc}
\end{table}

For each Q-learning episode: we sample one policy $\pi_i$ for player $i$ from the distribution $\pi^r_i$, and then the sampled policy $\pi_i$ and the opponent Q-learning agent for $\beta_{-i}$ play an episode against each other. If the sampled policy $\pi_i$ corresponds to the new strategy $\nu_i$, it plays with $\epsilon$-greedy exploration. The opponent Q-learning agent always plays with $\epsilon$-greedy exploration. The episode is used to update the Q-learning agent for $\beta_{-i}$. The episode is also used to update the Q-learning agent for $\nu_i$, regardless of whether or not the chosen $\pi_i$ is $\nu_i$.

\subsection{APSRO}

Tabular APSRO experiments were performed according to \cite{anytime_psro}. We use the same code as for SP-PSRO, with the difference being that we do not create a new policy $\nu_i$. 

\subsection{PSRO}

For tabular PSRO experiments: in each iteration, we first compute the empirical game payoff matrix, then use linear programming to find the Nash of the empirical game. We train Q-learning agents for each player against the other's empirical game Nash, and then add these to the population. The Q-learning agents use the same hyperparameters as for SP-PSRO and A-PSRO.

\section{Neural Training Details}

\subsection{PSRO}
For neural experiments, the PSRO empirical payoff matrix is estimated using 3000 evaluation rollouts per policy matchup, and the meta-game NE is calculated using 2000 iterations of Fictitious Play \citep{fp}. APSRO and SP-PSRO skip calculating the empirical payoff matrix. We do not count experience used to generate payoff matrix utilities in comparisons with PSRO.

\subsection{APSRO}
We use the same neural APSRO method as provided by \citet{anytime_psro}, using the Multiplicative Weights Update (MWU) algorithm as the no-regret metasolver with a learning rate of 0.1 and updating every 10th RL iteration. Action payoffs for MWU corresponding to expected utilities for population policies in $\Pi_i^t$ against the current $\beta_{-i}$ are estimated by averaging the empirical payoffs from the last 1000 rollouts in which each population policy was sampled. Exploitability is measured against the time-average of the MWU mixed-strategy from each APSRO iteration.

\subsection{SP-PSRO}

For SP-PSRO, we use the same MWU no-regret solver and parameters as we do with APSRO, where the actively-learning new strategy $\nu_{i}$ is included as an action for the no-regret solver. Because $\nu_{i}$ would by default only collect experience when the no-regret solver samples it, we additionally provide $\nu_{i}$ with off-policy experience from all other policies in the population $\Pi_i^t$ when they are sampled and generate experience as well.

We train the time-average $\bar{\nu_i}$ of $\nu_i$ as a neural network, and to do so, we save all experience generated by $\nu_i$ to a buffer using reservoir sampling \citep{vitter1985random, nfsp} with a maximum capacity of 2e6 samples. After BR training is complete, we use supervised learning to train a softmax policy on the reservoir buffer data with cross-entropy loss on actions given observations to distill the time-average of the new policy $\nu_i$. To ensure that enough experience from $\nu_i$ is always generated and added to the reservoir buffer, a small fixed portion $p$ of all experience rollouts in the BR training process is forced to be played as a matchup between $\nu_i$ and a non-exploring evaluation copy of $\beta_{-i}$. 

For Liars Dice, $p=0.05$, and for Small Battleship and 4x Repeated RPS, $p=0.1$. We train each $\bar{\nu_i}$ on the reservoir buffer data with a learning rate of 0.1 for 10,000 SGD batches. We use an MLP with three 128-unit layers and ReLu activations for $\bar{\nu_i}$ in all games.

Exploitability is measured against the time-average of the MWU mixed-strategy from each SP-PSRO iteration where $\bar{\nu_i}$ is used to represent the new strategy.

When the new population strategy $\nu_i$ and the BR $\beta_{-i}$ collect experience against each other, unless otherwise stated, they both use and play against exploring $\epsilon$-greedy versions of each other.

\subsection{Best Responses}

Hyperparameters to train deep RL best responses for each game are provided below. We use DDQN \cite{van2016deep} to train RL best responses for all neural experiments. Any hyperparameters not listed are default values in RLlib \citep{liang2018rllib} version 1.0.1.

\begin{table}[H]
\centering
\begin{tabular}{ll}
algorithm & DDQN \\
circular replay buffer size & 50,000 \\
prioritized experience replay & No \\
total rollout experience gathered each iter & 2048 steps \\
learning rate & 0.0026 \\
batch size & 4096 \\
optimizer & Adam \citep{kingma2014adam} \\
TD-error loss type & MSE \\
target network update frequency & every iteration \\
MLP layer sizes & [128, 128] \\
activation function & ReLu \\
discount factor $\gamma$ & 1.0 \\
best response RL process stopping condition & 7.5e5 timesteps \\
exploration $\epsilon$ & Linearly annealed from 0.06 to 0.001 \\
& over 2e5 timesteps \\
\end{tabular}
\caption{Liar's Dice Deep RL Best Response Hyperparameters}
\label{table:ddqn-leduc}
\end{table}

\begin{table}[H]
\centering
\begin{tabular}{ll}
algorithm & DDQN \\
circular replay buffer size & 200,000 \\
prioritized experience replay & No \\
total rollout experience gathered each iter & 1024 steps \\
learning rate & 0.0019 \\
batch size & 2048 \\
optimizer & Adam \citep{kingma2014adam} \\
TD-error loss type & MSE \\
target network update frequency & every 1e5 timesteps \\
MLP layer sizes & [128, 128, 128] \\
activation function & ReLu \\
discount factor $\gamma$ & 1.0 \\
best response RL process stopping condition & 3e5 timesteps (Repeated RPS) \& \\
& 7.5e5 timesteps (Battleship) \\
exploration $\epsilon$ & Linearly annealed from 0.06 to 0.001 \\
& over 2e6 timesteps \\
\end{tabular}
\caption{4x Repeated RPS and Small Battleship Deep RL Best Response Hyperparameters}
\label{table:ddqn-rps_battleship}
\end{table}

\section{Computational Costs}
Experiments were run on local machine with 128 logical CPU cores, 4 Nvidia RTX 3090 GPUs, and 512GB of RAM. Each tabular experiment run used a single core, and each neural experiment run used up to 5 CPU cores per player to train best responses and up to 4 CPU cores to evaluate meta-game empirical payoffs, for a maximum total of 14 cores per neural experiment. All neural experiments individually used less than 5GB of VRAM. Tabular and neural experiments had durations between 1 and 7 days.

\section{Code}
We will include a GitHub link to our open source code under the MIT license soon. 
Our code is built on top of the OpenSpiel \citep{lanctot2019openspiel} and RLlib \citep{liang2018rllib} frameworks, both of which are open source and available under the Apache-2.0 license.

\end{document}